\documentclass[prd,showpacs,superscriptaddress,amsmath,twocolumn]{revtex4}
\newcommand{\paslash}{\ensuremath {\slash}\hspace{-0.2cm}\partial\/}
\newcommand{\paslashnew}{\ensuremath {\slash}\hspace{-0.23cm}\bar{P}\/}
\newcommand{\aslash}{\ensuremath {\slash}\hspace{-0.25cm}A\/}
\newcommand{\pislash}{\ensuremath {\slash}\hspace{-0.2cm}\Pi\/}
\usepackage{epsfig}

\begin{document}

\title{Color-flavor locked superconductor in a magnetic field}

\author{Jorge L. Noronha}
\email{noronha@fias.uni-frankfurt.de}
\affiliation{Frankfurt Institute for Advanced Studies, J.W. Goethe--Universit\"at, D-60438
Frankfurt am Main, Germany}

\author{Igor A. Shovkovy}
\email{I-Shovkovy@wiu.edu}
\affiliation{Department of Physics, Western Illinois University, Macomb, IL 61455, USA}

\begin{abstract}
We study the effects of moderately strong magnetic fields on the
properties of color-flavor locked color superconducting quark matter
in the framework of the Nambu-Jona-Lasinio model. We find that the
energy gaps, which describe the color superconducting pairing as well
as the magnetization, are oscillating functions of the magnetic field.
Also, we observe that the oscillations of the magnetization can be
so strong that homogeneous quark matter becomes metastable for a
range of parameters. We suggest that this points to the possibility
of magnetic domains or other types of magnetic inhomogeneities in
the quark cores of magnetars. 
\end{abstract}

\date{\today}
\maketitle

\section{Introduction}

It has long been suggested that the superdense core of compact stars
may be composed of deconfined quark matter \cite{itoh}. However,
calculations using effective models for the strongly coupled limit of
quantum chromodynamics (QCD) \cite{cs,cs4fermi}, as well as ab initio
calculations performed in weak coupling \cite{cs-weak,properties},
predict that a phase transition between normal and color superconducting
(CSC) quark matter takes place at high baryon densities and sufficiently
low temperatures (for early papers on color superconductivity see
Ref.~\cite{love}, and for recent reviews see Ref.~\cite{reviews}).
Therefore, it is plausible that the inner region of compact stars may
consist of color superconducting matter.

Strong magnetic fields exist on the surface of compact stars, e.g.,
$B \lesssim 10^{12}$~G for ordinary neutron stars \cite{reviewNS},
while for magnetars they can be as large as $B \simeq 10^{16}$~G
\cite{thompson}. Also, recent studies have given support to the idea
that old neutron stars could have previously been magnetars with magnetic
fields that decayed over many years \cite{heyl,pons}. This then
suggests that magnetars could also have color superconducting cores.
Therefore, a detailed study of the effects of strong magnetic fields on
color superconductors may be very important to understand the physics
of magnetars.

According to the standard magnetar model by Duncan and Thompson \cite{duncan},
the energy bursts of soft gamma ray repeaters (SGR's) are caused by solid fractures
induced by strong magnetic fields in the crust of the star. Many properties of SGR's
are well described using the hadronic magnetar model \cite{woods} but there are still
some remaining issues, such as the quasi-periodicity of active phases, which require 
alternative ways of understanding them \cite{zhang}.

The physical upper limit for the magnetic field in a gravitationally bound star is
$B\simeq 10^{18}$~G \cite{reviewNS,CPL}, which is obtained by comparing the energy of magnetic
and gravitational fields. However, if quark stars are self-bound rather than gravitationally
bound objects this upper limit could be even higher. The typical energy scale defined
by these strong fields is of the same order of magnitude as the color superconducting
gap present in the quasiparticles energy spectrum. Thus, one expects that strong magnetic 
fields should affect the diquark pairing. The first step in this
direction was presented in Ref.~\cite{cristina}, where the effects of an extremely 
strong magnetic field on the pairing of a color-flavor locked (CFL) superconductor
\cite{CFL} were studied. The medium magnetic field in the CFL phase
is a linear combination of a gluon and the usual vacuum magnetic fields
\cite{rotated-mag-field}. This ``rotated" magnetic field is not subject to the Meissner
effect and, thus, freely penetrates the system.

In this paper we investigate the effects of a strong magnetic field (namely,
$eB/\mu^2\lesssim 1$ where $\mu$ is the baryon chemical potential of quarks)
on color-flavor locked superconducting quark matter. This is expected to model
the conditions which exist at the inner core of magnetars, where the density of matter
is up to ten times higher than the nuclear saturation density. For simplicity's sake, 
we assume that such matter is sufficiently dense so that effects of non-vanishing 
quark masses (including the strange quark mass) could be neglected. Also, using the
property of enforced neutrality of the color-flavor locked phase \cite{enforce_n},
we assume that no electrons are needed in the magnetized version of the phase.
By applying the same arguments as in Ref.~\cite{enforce_n}, we expect that
this is justified as soon as all quark quasiparticles remain gapped. In our analysis,
therefore, the chemical potentials of up, down and strange quarks are equal. 
Admittedly, we neglect possible non-zero ``color" chemical potentials that
might be needed to enforce the color Gauss law \cite{BubSho}. However, such chemical potentials
are expected to be small for the range of parameters studied in this paper, i.e., 
$eB/\mu^2\lesssim 1$.

In order to determine the thermodynamic properties of the magnetic CFL (mCFL)
phase, we solve the corresponding gap equations for a range of magnetic fields.
As expected, the magnetization of quark matter displays de Haas-van Alphen
oscillations and it can be as large as the applied magnetic field for a wide
range of parameters. This is in contrast to the magnetization of dense hadronic
matter, which is usually considered to be negligible \cite{lattimer}. The large magnetization
in magnetar models with (either normal or color superconducting) quark matter
cores could be related to physical properties that are distinctively different from
models with only hadronic matter.

In the next section we discuss the most important features concerning mCFL color 
superconductors and also illustrate the derivation of the free energy and the corresponding 
gap equations (details of our calculations are explicitly given in Appendix~A and B). 
In the same section we discuss the regularization scheme employed in our 
calculations and, in Sec.~\ref{sectionIII}, we show our numerical results. In the last
section we give our conclusions. Throughout the paper the units are $\hbar=c=k_{B} = 1$
and 4-vectors are denoted by capital letters, e.g., $K^{\mu} = (k_{0},\vec{k})$.
Also, the standard Minkowski metric $\eta_{\mu\nu}={\rm diag}(+,-,-,-)$ is used.

\section{The free energy of the magnetic CFL phase}
\label{sectionII}

In cold, dense quark matter, color superconductivity is expected to occur due to
attractive interactions between quarks located on the Fermi surface. At very low
temperatures and extreme high densities one can safely neglect the effects of
nonzero quark masses and, in this case, the CFL phase is expected to be the true
ground state of quark matter \cite{CFL}. This phase is characterized by the condensation
of quark Cooper pairs in the color-antitriplet, flavor-antitriplet representation, which breaks
the initial symmetry group $[SU(3)]_{C}\otimes SU(3)_{L}\otimes SU(3)_{R}\otimes
U(1)_{B}\otimes U(1)_{A}$ into the diagonal $SU(3)_{R+L+C}$ subgroup (note that
$U(1)_{A}$ is an approximate symmetry in dense matter). After taking into account
that all the gluons become massive through the Higgs mechanism, one finds that ten
Nambu-Goldstone bosons appear due to the breaking of global symmetries
\cite{effCFL,sonstephanov}. Following the same convention as in the QCD vacuum, 
these bosons are the $\pi^{\pm}$, $\pi^0$, $K^{\pm}$, $K^0$, $\bar{K}^0$, $\eta$, and $\eta^\prime$
mesons, and $\varphi$, which is a massless field related to the baryon symmetry breaking.

The CFL phase has some very unusual properties, for instance, it is a baryon superfluid
but not an electromagnetic superconductor \cite{CFL}. In fact, when it comes to its
electromagnetic properties a CFL superconductor is more adequately described as
an isotropic electromagnetic insulator \cite{enforce_n}. In conventional superconductivity,
the $U(1)$ gauge symmetry of electromagnetism is spontaneously broken by the
condensation of electron Cooper pairs, which gives an effective mass to the photon
that screens external magnetic fields (electromagnetic Meissner effect). On the other
hand, in color superconductors the initial $[U(1)]_{em}$ symmetry, whose generator
coincides with one of the vector-like generators of $SU(3)_{L}\otimes SU(3)_{R}$, is
not really broken but ``rotated" \cite{CFL}. The new group, denoted here as
$[\widetilde{U}(1)]_{em}$, corresponds to a massless linear combination of the
vacuum photon and the eighth gluon. This is analogous to the electromagnetic
$[U(1)]_{em}$ symmetry that remains unbroken after the electroweak symmetry
breaking. Using the same convention for colors as in Ref.~\cite{cristina,cristinalong},
it can be shown \cite{CFL,Litim} that the ``rotated"  electric $\widetilde{Q}$-charges
of the quarks are
\begin{equation}
\begin{tabular}{|c|c|c|c|c|c|c|c|c|}
  \hline
  $s_{b}$ & $s_{g}$ & $s_{r}$ & $d_{b}$ & $d_{g}$ & $d_{r}$ & $u_{b}$ & $u_{g}$ & $u_{r}$ \\
  \hline
  0 & 0 & - & 0 & 0 & - & + & + & 0 \\
  \hline
\end{tabular}   \nonumber
\end{equation}
in units of the $\widetilde{Q}$-charge of the electron, which is $\widetilde{e} = e \cos{\theta}$
with $\theta\equiv \arccos(g/\sqrt{g^2+e^{2}/3})$ as the mixing angle \cite{rotated-mag-field}.
In color-flavor space, the associated charge operator $\widetilde{Q}$ is given by
$\widetilde{Q}=Q_{f}\otimes \openone_{c}-\openone_{f}\otimes Q_{c}$,
where $Q_{c}=-\lambda_{8}/\sqrt{3}={\rm diag}(-1/3,-1/3,2/3)$.

The presence of an external magnetic field already reduces the symmetry of the model
because only $s$ and $d$ quarks have the same electric charges, i.e., $SU(3)_{L}\otimes
SU(3)_{R}\to SU(2)_{L}\otimes SU(2)_{R}$. Therefore, the pattern of symmetry breaking
that defines the mCFL phase is given by $[SU(3)]_{C}\otimes SU(2)_{L}\otimes SU(2)_{R}
\otimes U(1)_{B}\otimes U(1)_{A} \otimes U(1)^{-}_{A} \Longrightarrow SU(2)_{R+L+C}$,
where $U(1)^{-}_{A}$ corresponds to an anomaly-free current formed by a linear
combination of the $s$, $d$, and $u$ axial currents \cite{igoraxial}. Note that in
this case there are only six Nambu-Goldstone bosons (i.e., $\pi^0$, $K^0$,
$\bar{K}^0$, $\eta$, $\eta^\prime$, and $\varphi$) due to symmetry
breaking and all of them are neutral with respect to the $\widetilde{Q}$-charge.
As expected, the masses of the charged Nambu-Goldstone bosons of the 
CFL phase (i.e. $\pi^{\pm}$ and $K^{\pm}$) have terms that are proportional 
to $\sqrt{eB}$ \cite{ferrerlowenergy} (see also Ref.~\cite{manuel-eff}).

It is convenient at this point to discuss the energy scales that appear in our study in
more detail. The relevant mass scales in the problem are the quark chemical potential
$\mu$, the temperature $T$, and the magnetic length $l=1/\sqrt{eB}$. For very high,
but still realistic densities, $\mu\simeq 0.5$~GeV. If the magnetar's interior fields are 
indeed as large as $B\lesssim 10^{18}$~G the corresponding magnetic energy scale approaches
the QCD scale, $\sqrt{eB}\lesssim 77$~MeV. (Note the following relation: $\sqrt{eB} \simeq 
7.69\times 10^{-8} \sqrt{B/1\,\mbox{G}} $~MeV.) The relevant dimensionless parameter in
the problem is the ratio $eB/\mu^2$, which is smaller than one even for
the largest possible fields. Thus, from a phenomenological point of view it would suffice
to study only the regime of small $eB/\mu^2$. However, our analysis extends also to larger 
fields that correspond to $eB/\mu^2 \simeq 1$. 
Note that at such extremely high fields, the magnetic length is comparable or even
smaller than the average interquark distance. Then, all usual assumptions about 
the color superconductor's in-medium electromagnetic properties should be 
reanalyzed. In fact, in this limit the most important effect caused by the field might
be the so-called magnetic catalysis of chiral symmetry breaking, as in the vacuum 
\cite{igoraxial,catalysis,catalysis-others}. The corresponding new ground state is
characterized by different condensates. Its symmetry, though, is similar to the
symmetry in the mCFL phase, except for the baryon number symmetry that remains
unbroken \cite{igoraxial}.

We use a massless three-flavor quark model with a local NJL-type interaction to describe
the mCFL phase. The free energy density $\mathcal{F}$ of this system composed of
quarks in the presence of an external magnetic field ${H}$ is a functional that depends
on the gap functions, the chemical potential, the temperature, and the induced magnetic field ${B}$.
These two fields have different physical meaning and need to be distinguished. 
For instance, in an idealized model for the core of a color superconducting magnetar, 
${H}$ corresponds to the field present in the outer layers of the star. Inside the 
core one has to take into account the magnetization of the medium and in this 
case ${B}$ has to be used. We assume that both ${H}$ and ${B}$ are uniform fields
that point in the $\hat{z}$-direction. This approximation is valid as long as the fields do
not change appreciably within the relevant length scales defined throughout the computation
of the system's thermodynamic quantities such as the quark number density. This is
indeed the case here because the characteristic length scale associated with the problem
is of the order of one fermi.

The partition function of the system is given by
\begin{eqnarray}
Z &=&\mathcal{N}\,e^{-\beta V \left(\mathcal{F}+\frac{B^2}{8\pi}\right)}
\nonumber\\
&=&\int\,\mathcal{D}\bar{\psi}\mathcal{D}\psi\,\exp{\left\{\int d^{4}X\,
\left[\mathcal{L}-\frac{B^2}{8\pi}\right]\right\}},
\label{partitionfunction}
\end{eqnarray}
where $\mathcal{N}$ is the normalization constant. The Lagrangian density reads
\begin{equation}
\mathcal{L}=\bar{\psi}(i\paslash+e\widetilde{Q}\,\aslash+\mu\gamma_{0})\psi
+\sum_{\eta=1}^{3}\frac{G}{4}(\bar{\psi}P_{\eta}\psi_{c})(\bar{\psi}_{c}\bar{P}_{\eta}\psi),
\label{lagrangian}
\end{equation}
where $A^{\mu}$ describes the gauge field of the in-medium electromagnetism
$[\widetilde{U}(1)]_{em}$, $\beta=1/T$ is the inverse temperature, and
$V$ is the 3-volume. The quark spinor $\psi_{\alpha}^{a}$ carries color
$a=(b,g,r)=(1,2,3)$ and flavor $\alpha=(s,d,u)$ indices, and the
charge-conjugate spinors are defined as $\psi_{c}=C\bar{\psi}^{T}$ and
$\bar{\psi}_{c}=\psi^{T}C$, where $C=i\gamma^{2}\gamma^{0}$ is the charge
conjugation matrix. We consider only the pairing in the antisymmetric
channels and, thus, $(P_{\eta})_{\alpha\beta}^{ab}=i\gamma_{5}\,
\epsilon^{ab\eta}\epsilon_{\alpha\beta\eta}$ (no sum over $\eta$)
and $\bar{P}_{\eta}=\gamma_{0}P_{\eta}^{\dagger}\gamma_{0}$.
The index $\eta$ labels the pairing channels,  i.e., $\eta=1$, $2$, and $3$
correspond to $ud$, $us$, and $sd$ pairings, respectively.

For each channel we introduce a complex scalar field $\phi_{\eta}$,
with expectation value $\Delta_{\eta}$. The four-fermion interaction is
bosonized via a Hubbard-Stratonovich transformation, which then gives a
Yukawa-type interaction,
\begin{equation}
\frac{G}{4}(\bar{\psi}P_{\eta}\psi_{c})(\bar{\psi}_{c}\bar{P}_{\eta}\psi)
\to \frac{\phi_{\eta}}{2}(\bar{\psi}_{c}\bar{P}_{\eta}\psi)
+\frac{\phi^{*}_{\eta}}{2}(\bar{\psi}P_{\eta}\psi_{c})-\frac{|\phi_{\eta}|^2}{G}.
\label{interaction}
\end{equation}
In the following, we neglect diquark fluctuations and, thus, set
$\phi_{\eta}=\Delta_{\eta}$. Moreover, using the standard Nambu-Gorkov spinor
\begin{equation}
\Psi=
\begin{pmatrix}
  \psi  \\
  \psi_{c} \\
\end{pmatrix},
\label{nambugorkov}
\end{equation}
and the gap matrix $\Phi^{+}=\sum_{\eta=1}^{3}\Delta_{\eta}P_{\eta}$,
the Lagrangian density can be rewritten as
\begin{equation}
\mathcal{L}(X)=-\sum_{\eta=1}^{3}\frac{|\Delta_{\eta}|^2}{G}
+\frac{1}{2}\bar{\Psi}(X)\,\mathcal{S}^{-1}(X)\Psi(X)\, ,
\label{lagrangian1}
\end{equation}
where
\begin{equation}
\mathcal{S}^{-1}(X) =
\begin{pmatrix}
  [G_{0(\widetilde{Q})}^{+}]^{-1} & \Phi^{-}  \\
  \Phi^{+} &[G_{0(-\widetilde{Q})}^{-}]^{-1}\\
\end{pmatrix}\, ,
\label{def_S-11}
\end{equation}
and the following shorthand notation is used:
$[G_{0(\widetilde{Q})}^{\pm}]^{-1}
=[i\paslash+e\widetilde{Q}\,\aslash(X)\pm\mu\gamma_{0}]$
and $\Phi^{-}=\gamma_{0}(\Phi^{+})^{\dagger}\gamma_{0}$.

The mCFL pairing is characterized by the residual $SU(2)_{R+L+C}$ global symmetry, which
means that the corresponding gap matrix is invariant under simultaneous flavor ($1\leftrightarrow 2$)
and color ($1\leftrightarrow 2$) exchanges. This implies that $\Delta\equiv\Delta_{1}=\Delta_{2}$ and
$\phi\equiv\Delta_{3}$. The CFL gap structure is recovered when $\Delta=\phi$.

Even though all diquark pairs are chargeless with respect to the rotated 
electromagnetism they can be composed of either quarks with opposite $\widetilde{Q}$-charges or
$\widetilde{Q}$-neutral quarks (in the following, when discussing quark charges
we always have in mind the $\widetilde{Q}$-charges even if this is not explicitly 
emphasized). The gap function $\phi$ has only contributions from pairs of neutral quarks,
while $\Delta$ is formed by pairs of quarks with opposite charges and pairs of neutral
quarks. In the presence of a strong magnetic field one would naively expect that pairs
made of quarks with opposite charges have smaller coherence lengths in comparison to the pairs formed
only by neutral quarks, i.e., $\Delta$ should be larger than $\phi$. In the next section
we show that this is the case in the limit of very large fields, but is not generally true.

The Gibbs free energy density $\mathcal{G}$ of the mCFL phase is
\begin{equation}
\mathcal{G}=\frac{B^2}{8\pi}-\frac{HB}{4\pi}+\mathcal{F}-\mathcal{F}_{\rm vac}\, ,
\label{partitionfunction1}
\end{equation}
where
\begin{equation}
\mathcal{F}=\frac{2\Delta^2}{G}+\frac{\phi^2}{G}-\Gamma(T,\mu,\Delta,\phi,B).
\label{F_function1}
\end{equation}
The last term is the one-loop contribution of quarks, i.e.,
\begin{equation}
\Gamma(T,\mu,\Delta,\phi,B)=\frac{1}{2}\ln \det\mathcal{S}^{-1}.
\label{Gamma}
\end{equation}
The free energy of the vacuum is given by
$\mathcal{F}_{\rm vac}=-\Gamma_{\rm vac} \equiv-\Gamma(0,0,0,0,0)$.

In equilibrium $\mathcal{G}$ is evaluated at its stationary point with
respect to $\Delta$, $\phi$, and $B$ and it describes all the system's
thermodynamic properties. The CFL free energy, $\mathcal{F}_{0}$, 
has to be recovered when $H \to 0$. The stationary
point corresponds to the solutions of the equations
\begin{subequations}
\begin{eqnarray}
\Delta &=&\frac{G}{4}\Big(\frac{\partial \Gamma}{\partial \Delta}\Big),
\label{stationaryequations1}\\
\phi &=& \frac{G}{2}\Big(\frac{\partial \Gamma}{\partial \phi}\Big),
\label{stationaryequations2}\\
B&=&H+4\pi M,
\label{stationaryequations3}
\end{eqnarray}
\end{subequations}
where $M$ is the magnetization of the system, which is given by
$M=(\partial \Gamma/\partial B)|_{\rm stationary}$. At the 
stationary point, the gaps depend explicitly on the induced
field $B$. This field incorporates the magnetic properties of the medium,
described by the magnetization $M$.

In this paper we consider the zero temperature case only, i.e., $T=0$ 
because typical temperatures of matter in stellar cores are much less than $1$~MeV
\cite{pons}. According to Eq.~(\ref{thermofunctionmCFLtemp}) in Appendix~\ref{appA},
the one-loop quark contribution to the free energy reads
\begin{eqnarray}
\Gamma(0,\mu,\Delta,\phi,B)&=&3\,P(\phi)+P(\Delta_{1})+P(\Delta_{2})\nonumber \\ &+& 4\,F(\Delta),
\label{thermofunctionmCFL}
\end{eqnarray}
where $\Delta_{1/2}=\frac{1}{2}(\sqrt{\phi^2+8\Delta^2}\pm\phi)$.
The neutral quark contributions are given by the three
terms containing $P$-functions
\begin{eqnarray}
 P(\phi) &=& \mbox{tr}_{B=0}\left[E_{0}^{+}(\phi)+E_{0}^{-}(\phi)\right],
\label{functionPzero}
\end{eqnarray}
where the trace is defined as
$\mbox{tr}_{B=0}[\ldots]=\int\frac{d^{3}\vec{p}}{(2\pi)^3}[\ldots]$,
and $E_{0}^{\pm}[\phi]=\sqrt{(p \mp \mu)^2 +\phi^2}$
with $p\equiv \sqrt{p_{3}^2+\vec{p}_{\perp}^{\,2}}$. On the other hand, the 
one-loop contributions that come from charged quarks are given in terms of
the following function:
\begin{eqnarray}
 F(\Delta) &=& \mbox{tr}_{B}\left[E_{B}^{+}(\Delta)+E_{B}^{-}(\Delta)\right].
\label{functionFzero}
\end{eqnarray}
The trace here is defined as a sum over the Landau levels with an integral over
the longitudinal momentum, 
$\mbox{tr}_{B}[\ldots]=\frac{eB}{8\pi^2}\sum_{n=0}^{\infty}\alpha_{n}\,
\int_{-\infty}^{\infty}dp_{3}[\ldots]$ with $\alpha_{n}=2-\delta_{n0}$. 
The quasiparticle dispersion relation of charged quarks
is $E_{B}^{\pm}[\Delta]=\sqrt{(\varepsilon_{B} \mp \mu)^2 +\Delta^2}$, 
where $\varepsilon_{B}=\sqrt{p_{3}^2+2eBn}$. 

Using the well-known Euler-McLaurin summation formula
\begin{equation}
\sum_{n=0}^{\infty}\frac{\alpha_{n}}{2}\,f(n)
=\int_{0}^{\infty}dx\,f(x)-\frac{1}{12}f'(0)+\cdots ,
\label{serieszerofield}
\end{equation}
we see that, formally, $\lim_{B\to 0}F(x)=P(x)$. Therefore, after taking
the limit $\Delta=\phi$ and $B=H=0$ we recover the free energy of the CFL phase
in the absence of an external field.

The traces in the definition of the $P$ and $F$ functions involve integrations 
and sums over the whole phase space, which means that these functions 
diverge in the ultraviolet. In order to obtain the physically meaningful free 
energy density of the system, these functions must be regularized. In 
effective quark models such as the Nambu-Jona-Lasinio model used here, 
it is common to restrict the phase space by introducing a finite sharp cutoff 
in momentum space. However, because of the special properties of the system 
in a magnetic field such a prescription is not very useful. Utilizing a 
sharp cutoff when an energy spectrum with discrete Landau levels is considered
would introduce unphysical discontinuities in many thermodynamical quantities. 

In this study, therefore, we regularize the traces in Eqs.~(\ref{functionPzero}) 
and (\ref{functionFzero}) by introducing a smooth cutoff function $h_{\Lambda}$ 
(where $\Lambda$ is a constant with the dimension of energy). The cutoff 
function should approach $1$ at small energies (i.e., $\varepsilon\ll \Lambda$) 
and $0$ at large energies (i.e., $\varepsilon\gg \Lambda$). Providing that 
$h_{\Lambda}$ falls off sufficiently fast in the ultraviolet, the regularized 
functions,
\begin{eqnarray}
 P_{\Lambda}(\phi) &=& \int\frac{d^{3}\vec{p}}{(2\pi)^3}
h_{\Lambda}\left[E_{0}^{+}(\phi)+E_{0}^{-}(\phi)\right],
\label{functionPzeroreg}
\end{eqnarray}
and
\begin{eqnarray}
 F_{\Lambda}(\Delta) &=& \frac{eB}{8\pi^2}\sum_{n=0}^{\infty}\alpha_{n}
\int_{-\infty}^{\infty}dp_{3}h_{\Lambda}\,
\left[E_{B}^{+}(\Delta)+ E_{B}^{-}(\Delta)\right].\nonumber \\
\label{functionFzeroreg}
\end{eqnarray}
are free from divergences.

After using several different choices for $h_{\Lambda}$ we decided on the 
Gaussian-like form, i.e., 
\begin{equation}
h_{\Lambda}=\exp\left(-\varepsilon^2 / \Lambda^2   \right),
\label{cutofffunction}
\end{equation}
where $\varepsilon=p$ in Eq.~(\ref{functionPzeroreg}) and $\varepsilon=\varepsilon_{B}$ 
in Eq.~(\ref{functionFzeroreg}). Note that a sharp cutoff could be implemented with the
function $h_{\Lambda}=\theta\left(\Lambda-\varepsilon\right)$ [where $\theta\left(x\right)$ 
is the step function]. While this would produce no apparent abnormalities in the behavior 
of $P_{\Lambda}(\phi)$, it leads to unphysical discontinuities in $F_{\Lambda}(\Delta)$ and,
consequently, in the free energy as a function of the magnetic field. 

In the numerical calculation in Sec.~\ref{sectionIII}, we use the cutoff function
in Eq.~(\ref{cutofffunction}) with $\Lambda=1$~GeV. It should be mentioned 
here that we have checked the cutoff independence of our qualitative results 
by varying the value of $\Lambda$ from $1$~GeV to $2$~GeV, and 
simultaneously readjusting the diquark coupling constant $G$ so as to keep 
the CFL gap at $B=0$ unchanged. To avoid a potential confusion, let us additionally 
mention that the dependence of the results on the cuttoff parameter $\Lambda$ 
cannot be completely eliminated in the analysis. This is a general feature of all 
non-renormalizable effective models. 

When using the smooth cutoff function (\ref{cutofffunction}) in our calculation, 
we get no divergences in the theory. Because no consistent continuum 
limit ($\Lambda\to \infty$) exists in the model at hand, there is also no need for 
any additional (finite) renormalization and/or subtractions in our calculations. 
Strictly speaking, one should only remember that all physical parameters of the 
model (e.g., the charge and the magnetic field) are defined at the scale of the 
cutoff (i.e., $1$~GeV for our choice of cutoff $\Lambda$). Because of the slow 
logarithmic running in the complete gauge theory, the corresponding values 
of parameters differ by less then $5\%$ from the parameters defined at the 
conventional infrared scale set by the electron mass. Also, since most of 
the results of interest depend on the renormalization group invariant quantity 
$eB$, we choose to ignore these minor corrections. [Alternatively, one might 
consider an explicit finite multiplicative renormalization $e_{\rm R}= e/\sqrt{1+C e^2}$ 
and $B_{\rm R}= B \sqrt{1+C e^2}$, with $C\propto \ln(\Lambda^2/m_e^2)$ \cite{Schwinger1951}. 
Here the new quantities, $e_{\rm R}$ and $B_{\rm R}$, are defined at the scale 
set by the electron mass. However, we do not find this necessary for the purposes 
of the current study.]

By making use of the regularized functions $P_{\Lambda}(\phi)$ and 
$F_{\Lambda}(\Delta)$, the gap equations (\ref{stationaryequations1}) 
and (\ref{stationaryequations2}) become
\begin{subequations}
\begin{eqnarray}
\Delta&=&\frac{G}{4}\left[R_{1} P_{1}(\Delta_{1})+R_{2} P_{1}(\Delta_{2})
                         +4\Delta F_{1}(\Delta)\right],
\label{gapequation1}\\
\phi&=&\frac{G}{2}\left[3\phi P_{1}(\phi)+U_{1} P_{1}(\Delta_{1})
                        +U_{2} P_{1}(\Delta_{2}) \right],
\label{gapequation2}
\end{eqnarray}
\end{subequations}
where the functions $P_{1}$ and $F_{1}$ are defined as follows
\begin{subequations}
\begin{eqnarray}
P_{1}(\phi)&\equiv& \frac{1}{\phi}\frac{d P_{\Lambda}(\phi)}{d\phi},
\label{derivfunctionP} \\
F_{1}(\Delta)&\equiv& \frac{1}{\Delta}\frac{d F_{\Lambda}(\Delta)}{d\Delta}.
\label{derivfunctionF}
\end{eqnarray}
\end{subequations}
In the gap equations above we also introduced the notation
\begin{subequations}
\begin{eqnarray}
R_{1/2}&=& 2\Delta\,\Big(1\pm\frac{\phi}{\sqrt{\phi^2+8\Delta^2}}\Big),
\label{functionsCandD1}\\
U_{1/2}&=& \pm\frac{\left(\phi\pm\sqrt{\phi^2+8\Delta^2}\right)^2}{4\sqrt{\phi^2+8\Delta^2}}.
\label{functionsCandD2}
\end{eqnarray}
\end{subequations}
The gap equations derived here are slightly different than those obtained in 
Refs.~\cite{cristina,cristinalong}. The reason is likely to be the difference in 
the structure of the interaction between the quarks. However, as shown in 
the next section, at asymptotically high fields our numerical solutions for 
the gap functions are in qualitative agreement with the analytical formulas 
obtained in Refs.~\cite{cristina,cristinalong}.

\section{numerical results}
\label{sectionIII}

\begin{figure}[!ht]
\epsfig{file=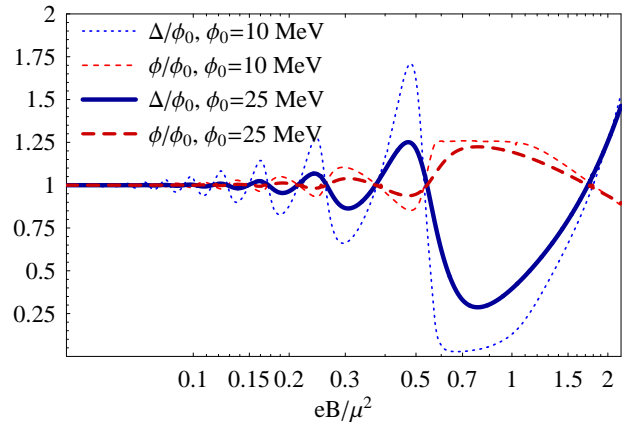,width=0.95\linewidth}
\caption{(Color online) Ratios $\Delta/\phi_{0}$ and $\phi/\phi_{0}$ 
versus $eB /\mu^2$ for two sets of parameters that yield
$\phi_{0}= 10$~MeV and $\phi_{0}= 25$~MeV.}
\label{figcompgaps}
\end{figure}

The free parameters of the model ($G$ and $\Lambda$) were set to yield a CFL 
gap of either $\phi_{0}= 10$~MeV or $\phi_{0}= 25$~MeV at $\mu=500$ MeV when 
$B=0$. (All numerical results below correspond to the following specific choice 
of the coupling constant and the cutoff parameter: $G=4.32~\mbox{GeV}^{-2}$ or
$G=5.15~\mbox{GeV}^{-2}$, and $\Lambda=1~\mbox{GeV}$.) In Fig.~\ref{figcompgaps} we 
plotted the ratio between the mCFL gaps and the CFL gap as functions of $eB/\mu^2$. 
As was mentioned in the last section, we have checked the robustness of our results 
by changing the parameters $\Lambda$ and $G$. When $eB/\mu^2 \gtrsim 0.3$ the 
gap $\Delta$, which receives contributions from pairs of chargeless quarks and also 
pairs with opposite charges, differs considerably from $\phi$, where only chargeless 
quarks enter in the pairing. For smaller fields ($eB/\mu^2 \lesssim 0.1$) the mCFL 
gaps are practically the same as $\phi_{0}$.

For ultrastrong fields ($eB/\mu^2 \gtrsim 2$) $\Delta$ is much larger 
than $\phi$. This is consistent with the analytical solutions of the mCFL gap 
equations found in Refs.~\cite{cristina,cristinalong}. There the authors considered 
fields so large that only the lowest Landau level contributes to the dynamics. 
As argued in the previous section, such fields should already probe the dynamics
of the vacuum and the dynamics due to the magnetic catalysis should be taken 
into account \cite{igoraxial,catalysis,catalysis-others}. Moreover, the corresponding 
field strengths appear to be of the order of $B\simeq 8.5\times 10^{19}$~G, 
assuming $\mu=500$ MeV, which may be too large to be found in compact stars 
\cite{lerche}.

In Fig.~\ref{figcompgaps}  we see that the gaps display magnetic oscillations with respect 
to $eB/\mu^2$. These oscillations share the same physical origin as the de Haas-van Alphen 
oscillations observed in metals. They appear as a consequence of the oscillatory structure 
in the density of states, which is imposed by the quantization of the energy levels associated 
the orbital motion of charged particles \cite{landau}. 

At zero temperature most electronic properties of metals depend on the density 
of states on the Fermi surface. Therefore an oscillatory behaviour as a function 
of $B$ should appear in any quantity that depends on the density of states on 
the Fermi energy. In the case of the mCFL color superconductor, it is the quark 
density of states on the Fermi surface that is relevant. Since the gaps in the 
excitation spectrum depend on the quark density of states and every physical 
quantity that we are interested in depends on the gaps, some type of magnetic 
oscillations should appear. The presence of nonzero gaps, however, smears out 
the Fermi surface so that the oscillatory structure is considerably reduced. 
This then explains the smoothness of the oscillations and their dependence on 
the magnitude of the gaps in Fig.~\ref{figcompgaps}.

In the presence of strong magnetic fields, the magnetization $M$ would affect 
the properties of the mCFL phase. As mentioned in the introduction, this might 
be relevant to explain the magnetic properties of magnetars with color superconducting 
quark cores. A phenomenon that is quite often observed in magnetic systems 
is the formation of magnetic domains \cite{landau}. As in metals, the large 
magnitude de Haas-van Alphen oscillations in the magnetization can create 
regions where $(\partial H/\partial B)_{\mu}<0$.
These correspond to unstable or metastable states. The condition for thermodynamic 
stability $(\partial H/\partial B)_{\mu}>0$ implies that $4\pi\,|\chi(\mu,B)| <1$, 
where $\chi (\mu,B)=(\partial M/\partial B)_{\mu}$ is the differential susceptibility. 
When the differential susceptibility exceeds $1/4\pi$, which depends on 
the geometry of the system, a transition into a magnetic domain configuration 
may occur. The presence of magnetic domains in the crust of a neutron star was 
discussed by Blandford and Hernquist in Ref.~\cite{blandford}.

\begin{figure}[!ht]
\epsfig{file=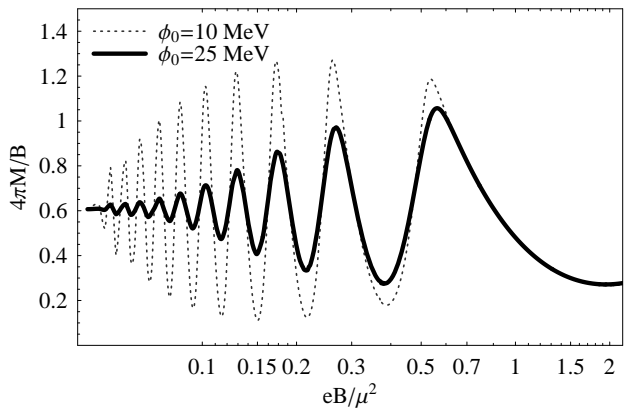,width=0.95\linewidth} 
\caption{Ratio $4\pi\,M/B$ versus $eB/ \mu^2$ for two sets of 
parameters that yield $\phi_{0}= 10$~MeV and $\phi_{0}= 25$~MeV.}
\label{figmag}
\end{figure}

In Ref.~\cite{lattimer} it was shown that the magnetization of hadronic matter is negligible 
even for magnetar conditions, i.e., $4\pi\, M/B\ll 1$ for $B\lesssim 10^{19} G$. In the case of 
color superconducting quark matter, however, the situation is very different. In Fig.~\ref{figmag} 
we plotted this ratio versus $eB/\mu^2$ for a mCFL superconductor. The magnetization is 
significantly larger in this case and it displays de Haas-van Alphen oscillations that have 
a very large magnitude. Note that nonzero energy gaps make the Fermi surface look fuzzy and the 
magnetization's oscillations in mCFL quark matter are much smoother than those shown 
by the magnetization of cold, unpaired quark matter \cite{ebert}.

\begin{figure}[!ht]
\epsfig{file=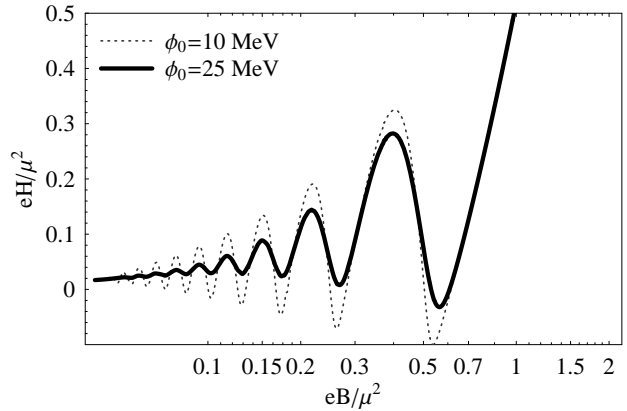,width=0.95\linewidth}
\caption{External field $eH/\mu^2$ versus $eB/\mu^2$ for two sets of 
parameters that yield $\phi_{0}= 10$~MeV and $\phi_{0}= 25$~MeV.}
\label{figmagGaussian}
\end{figure}

Our results for the $H$-$B$ curve are shown in Fig.~\ref{figmagGaussian}. Several regions 
of thermodynamic instability are obtained for $eB/\mu^2 \lesssim 1$. The metastable regions in 
Fig.~\ref{figmagGaussian}, which correspond to $(\partial H/\partial B)_{\mu}<0$, can be 
filtered out by either using a Maxwell construction or a mixed phase where microscopic 
domains with nonequal magnetizations coexist. This could lead to several physical
possibilities. As the field $H$ increases the system could undergo successive phase 
transitions with discontinuous changes of the induced magnetic field $B$ \cite{landau}.
When the mixed phases are formed, the relative size of domains with different 
magnetizations would change with $H$ so as to keep the average induced magnetic 
field $B$ continuous. In either case, since the magnitude of the fields involved is 
enormous, the system could potentially release an immense amount of energy. 
Further studies are needed to see whether this finding could have any important 
implications for magnetars.

\section{Conclusions}

In this paper we studied the effects of a moderately strong magnetic field 
on the Cooper pairing dynamics in cold and dense three-flavor quark matter. 
We solved the corresponding gap equations and calculated the magnetization 
for a wide range of magnetic fields, $eB/\mu^2\lesssim 1$. We found that, 
as the magnetic field increases, the system undergoes a continuous crossover 
from the usual CFL phase to the mCFL phase. Notably, for $eB/\mu^2 \lesssim 0.1$, 
which corresponds to $B\lesssim 4.2\times 10^{18}$ G (provided that $\mu=500$~MeV), 
there is no large difference between the mCFL and CFL gaps. 

In this study we also showed that the gaps display magnetic oscillations, 
which is a direct consequence of the Landau quantization of the energy 
levels in a magnetic field. Similar magnetic oscillations were predicted 
and later observed in type-II electronic superconductors \cite{yasui}. 
The effects these oscillations have on the transport properties of mCFL 
superconductors still remain to be understood.

We showed that the magnetization of mCFL quark matter displays de Haas-van 
Alphen oscillations whose amplitude can be as large as the magnetic field for a 
wide range of magnetic fields. Since the magnetization of dense hadronic matter 
is negligible, the nonzero magnetization of color superconducting quark matter 
may provide new observable effects that can help to distinguish purely hadronic 
magnetars from color superconducting ones. Our results for the oscillations of 
the magnetization suggest that homogeneous quark matter may become 
metastable for the range of parameters that are phenomenologically relevant to 
magnetars. Because of this there is a possibility that magnetic domains or other 
magnetic inhomogeneities can be found in the quark cores of magnetars. 
The successive phase transitions coming from discontinuous changes of the 
induced magnetic field $B$ during the stellar evolution can release a vast 
amount of energy that would heat up the star, which would then cool down 
by for example the emission of neutrinos. Therefore, bursts of neutrinos 
coming from magnetars with color superconducting cores even after the 
deleptonization period could be expected.

It has been recently pointed out in Ref.~\cite{ferrerchromo} that the 
chromomagnetic instability \cite{igorchromo}, which is present in neutral 
two-flavor color superconductivity at moderate densities \cite{g2sc}, can be 
removed by the formation of an inhomogeneous condensate of charged gluons and
the corresponding induction of a magnetic field. The field strengths used 
in their approach are comparable to those that cause the large magnitude
de Haas-van Alphen oscillations. Both effects and/or their competition may 
be relevant to the explanation of strong magnetic fields and other unusual 
observed properties of magnetars. 

There are several questions regarding color superconducting matter in 
strong fields and its consequent use in the study of superconducting 
magnetars that remain to be answered. One could, for example, study 
the cooling of their ultramagnetized cores. Also, the effects of strong 
magnetic fields on the global structure of the star, such as its mass and 
radius, are also very important and will be discussed elsewhere \cite{future}. 
The consideration of the effects of a nonzero strange quark mass on the 
low-energy description of mCFL color superconductors should also be 
investigated. 

The low-energy effective theory for the Nambu-Goldstone bosons in the mCFL 
phase was recently derived by Ferrer and Incera in Ref.~\cite{ferrerlowenergy}, 
although the pion decay constant and the meson maximum velocities still have to 
be determined within the microscopic theory. Using the zero field values of these 
parameters calculated in Ref.~\cite{sonstephanov} they obtained that the charged 
mesons decouple from the low-energy theory only when $eB \gtrsim 12\,\phi_{0}^2$. 
Incidentally, this corresponds to the region where the oscillations of the magnetization
and the gaps become already noticeable. This brings up an interesting question 
regarding the possible effects of the oscillations on the low-energy dynamics in
the mCFL phase. This problem is left for future study.

{\it Note added.} While finishing our paper we learned that a partially overlapping
study was done by Kenji Fukushima and Harmen J. Warringa \cite{FukWar}.

\section*{Acknowledgments} 

The authors thank E.~Ferrer, V.~Incera, O.~Kiriyama, C.~Manuel, D.~H.~Rischke, 
B.~Sa'd, and J.~Schaffner-Bielich for insightful discussions. J.L.N. acknowledges 
support by the Frankfurt International Graduate School for Science (FIGSS).


\appendix
\section{Evaluation of the determinant}
\label{appA}

In this appendix we present the details of the computation of 
$\Gamma(T,\mu,\Delta,\phi,B)$. The usual way to compute these 
determinants is to use the identity
\begin{equation}
\det
\begin{pmatrix}
  A & B  \\
  C & D \\
\end{pmatrix}
=\det\,(AD-ACA^{-1}B)
\label{det}
\end{equation}
for the determinant of a block matrix. Using this identity 
we can rewrite Eq.~(\ref{Gamma}) as
\begin{eqnarray}
\Gamma &=& \frac{1}{2}\ln \det\left\{[G_{0(\widetilde{Q})}^{+}]^{-1}
[G_{0(-\widetilde{Q})}^{-}]^{-1}\right.\nonumber\\
&&\left.-[G_{0(\widetilde{Q})}^{+}]^{-1}\Phi^{+}
[G_{0(\widetilde{Q})}^{+}]\Phi^{-}\right\},
\label{Gamma1}
\end{eqnarray}
where $[G_{0(\widetilde{Q})}^{\pm}]^{-1}=[i\paslash+e\widetilde{Q}\,\aslash(X)\pm\mu\gamma_{0}]$. 
However, the nontrivial color-flavor structure of $[G_{0(\pm\widetilde{Q})}^{\pm}]^{-1}$ complicates 
the calculations. 

\subsection{Introducing the charge projectors}

The easiest way to calculate Eq.~(\ref{Gamma1}) is to go back to the Lagrangian density and introduce the charge projectors in color-flavor space \cite{cristina,cristinalong}
\begin{subequations}
\begin{eqnarray}
\Omega_{(0)} &=& {\rm diag}(1,1,0,1,1,0,0,0,1),
\label{projector0}
\\
\Omega_{(+)} &=& {\rm diag}(0,0,0,0,0,0,1,1,0),
\label{projectorplus}
\\
\Omega_{(-)} &=& {\rm diag}(0,0,1,0,0,1,0,0,0),
\label{projectorminus}
\end{eqnarray}
\end{subequations}
which satisfy
\begin{equation}
\Omega_{(a)}\Omega_{(b)}=\delta_{ab}\Omega_{(b)},\quad a,b=0,+,-,\quad
\sum_{a=0,\pm}\Omega_{(a)}=1.
\label{projectorsproperties}
\end{equation}
In terms of these projectors the charge operator in color-flavor space 
is $\widetilde{Q}=\Omega_{(+)}-\Omega_{(-)}$. However, it is convenient 
to define the following charge operator in Nambu-Gorkov space 
$\widetilde{Q}_{NG}={\rm diag}_{NG}(\widetilde{Q},-\widetilde{Q})$, 
with eigenstates
\begin{equation}
\widetilde{Q}_{NG}\,\Psi_{(a)}=a\,\Psi_{(a)},
\label{nambugorkovchargedeigen}
\end{equation}
where
\begin{equation}
\Psi_{(a)}=
\begin{pmatrix}
  \psi_{(a)}  \\
  \psi_{c\,(-a)} \\
\end{pmatrix}, \quad
\psi_{(a)}=\Omega_{(a)}\psi.
\label{nambugorkovcharged}
\end{equation}
The charge operators can also be generalized to include the Nambu-Gorkov 
structure present throughout this derivation. In fact, one has
\begin{equation}
\Omega_{(a)}^{NG}=
\begin{pmatrix}
  \Omega_{(a)} & 0  \\
  0 & \Omega_{(-a)} \\
\end{pmatrix},
\label{nambugorkovchargedprojectors}
\end{equation}
which then satisfies
\begin{equation}
\Omega_{(a)}^{NG}\,\Omega_{(b)}^{NG}=\delta_{ab}\,\Omega_{(b)}^{NG},\quad a,b=0,+,-,\quad
\sum_{a=0,\pm}\Omega_{(a)}^{NG}=1,
\label{projectorspropertiesNG}
\end{equation}
where $\Psi_{(a)}=\Omega_{(a)}^{NG}\,\Psi$. Using these projectors Eq.~(\ref{lagrangian1}) can be expressed as
\begin{equation}
\mathcal{L} =-\sum_{\eta=1}^{3}\frac{|\Delta_{\eta}|^2}{G} + \sum_{a,b=0,\pm}\frac{1}{2}\bar{\Psi}_{(a)}
\Omega_{(a)}^{NG}\mathcal{S}^{-1}\Omega_{(b)}^{NG}\,\Psi_{(b)}.
\label{lagrangian2}
\end{equation}
However, it can be shown that
\begin{equation}
\Omega_{(a)}^{NG}\,\mathcal{S}^{-1}\Omega_{(b)}^{NG}=\delta_{ab}\,\Omega_{(a)}^{NG}\,\mathcal{S}^{-1}\Omega_{(a)}^{NG},
\label{inversepropagatorchargedproperties}
\end{equation}
where we used that $\Omega_{a}\Phi^{+}\Omega_{-a}$ is the only combination that is not identically zero. We define $\mathcal{S}_{(a)}^{-1}\equiv\Omega_{(a)}^{NG}\,\mathcal{S}^{-1}(X)$, which then implies that
\begin{equation}
\mathcal{S}^{-1}=\sum_{a=0,\pm}\Omega_{(a)}^{NG}\,\mathcal{S}_{(a)}^{-1}.
\label{inversepropagatorchargedproperties1}
\end{equation}
It is possible to show that $[\Omega_{(a)}^{NG},\mathcal{S}_{(b)}^{-1}]=0$.

The term involving the full inverse propagator in the Lagrangian density can be written as
\begin{eqnarray}
\frac{1}{2}\bar{\Psi}\,\mathcal{S}^{-1}\,\Psi &=& \frac{1}{2}\sum_{a=0,\pm}\bar{\Psi}_{(a)}\,\mathcal{S}^{-1}\,\Psi_{(a)}
\nonumber\\
&=&\frac{1}{2}\sum_{a=0,\pm}\bar{\Psi}_{(a)}\,\mathcal{S}_{(a)}^{-1}\,\Psi_{(a)}.
\label{inversepropagatorchargedproperties2}
\end{eqnarray}
Therefore, the Lagrangian density in Eq.~(\ref{lagrangian1}) can be rewritten as
\begin{equation}
\mathcal{L}=-\sum_{\eta=1}^{3}\frac{|\Delta_{\eta}|^2}{G}+\sum_{a=0,\pm}\frac{1}{2}\bar{\Psi}_{(a)}\mathcal{S}_{(a)}^{-1}\Psi_{(a)}
\label{lagrangian3}
\end{equation}
where
\begin{equation}
\mathcal{S}_{(a)}^{-1}=
\begin{pmatrix}
  [G_{0(a)}^{+}]^{-1} & \Phi_{(a)}^{-}  \\
  \Phi_{(a)}^{+} & [G_{0(a)}^{-}]^{-1} \\
\end{pmatrix}
\label{def_S-12}
\end{equation}
$[G_{0(a)}^{\pm}]^{-1}=\,\pislash_{(a)}\pm \mu\gamma_{0}$, and $\,\pislash_{(a)}=i\paslash+a\,e\,\aslash$.
The new gap matrices in the equation above are given by $\Phi_{(a)}^{+}=\Phi^{+}\Omega_{(a)}$ whereas
$\Phi_{(a)}^{-}=\gamma_{0}(\Phi_{(a)}^{+})^{\dagger}\gamma_{0}$.

\subsection{Lagrangian density in momentum space}

It is very convenient to express the Lagrangian density in Eq.~(\ref{lagrangian2})
in momentum space because it simplifies the computation of the determinants. Here we 
follow Ref.~\cite{cristinalong} and use the method originally developed for charged 
fermions by Ritus in Ref.~\cite{ritus}. In Ref.~\cite{ritusferrer}, the method was 
also extended to include charged vector fields. In this approach the diagonalization 
of the Green's functions of charged fermions in a uniform magnetic field in momentum 
space is obtained using the eigenfunction matrices $E_{p}(X)$. These functions are 
the wave functions corresponding to asymptotic states of charged fermions in a 
uniform background magnetic field.

The $E_{p}(X)$ functions were described in detail in Refs.~\cite{cristinalong,ritusferrer} 
and some of their main properties are presented in Appendix B. Using these eigenfunctions
it is possible to express the charged fields $\psi_{(\pm)}$ as
\begin{subequations}
\begin{eqnarray}
\psi_{(\pm)}(X) &=& \sum_{\bar{P}_{(\pm)}}E_{p}^{(\pm)}(X)\psi_{(\pm)}(\bar{P}_{(\pm)}),
\label{chargedfieldsmomentum}
\\
\bar{\psi}_{(\pm)}(X) &=& \sum_{\bar{P}_{(\pm)}}\bar{\psi}_{(\pm)}(\bar{P}_{(\pm)})\bar{E}_{p}^{(\pm)}(X),
\label{chargedfieldsmomentumdagger}
\end{eqnarray}
\end{subequations}
where, by definition, $\bar{E}_{p}^{(\pm)}(X)=\gamma_{0}\,(E_{p}^{(\pm)}(X))^{\dagger}\,\gamma_{0}$,
$\bar{P}_{(\pm)}=(p_0, 0, \pm\sqrt{2eBn},p_{3})$, and $n=0,1,2,\hdots$, denotes the
Landau levels. The 4-vector potential is in the Landau gauge,  i.e.,
$A^{\mu}=(0,0,Bx,0)$. Moreover, one can show that
\begin{equation}
 [G_{0(\pm)}^{\pm}]^{-1}(X)E_{p}^{(\pm)}(X)=E_{p}^{(\pm)}(X)[\,\paslashnew_{(\pm)}\pm \mu\gamma_{0}].
\label{inversepropfreemomentum}
\end{equation}
This implies that
\begin{widetext}
\begin{eqnarray}
\int d^{4}X\,\bar{\Psi}_{(a)}(X)\,\mathcal{S}_{(a)}^{-1}(X)\Psi_{(a)}(X)
&=&\sum_{\bar{P}_{(a)}}\bar{\Psi}_{(a)}(\bar{P}_{(a)})
\begin{pmatrix}
  [G_{0(a)}^{+}]^{-1}(\bar{P}_{(a)}) & \Phi_{(a)}^{-}  \\
  \Phi_{(a)}^{+} & [G_{0(a)}^{-}]^{-1}(\bar{P}_{(a)}) \\
\end{pmatrix}
\Psi_{(a)}(\bar{P}_{(a)}),
\label{lagrangianmomentumspace}
\end{eqnarray}
\end{widetext}
for $a=0,\pm$. Also, we have defined $\bar{P}_{(0)}=(p_0, \vec{p}\,)$ and
$[G_{0(0)}^{+}]^{-1}(\bar{P}_{(0)})=[\,\paslashnew_{(0)}\pm \mu\gamma_{0}]$.
In the following section we will compute the determinants using the momentum
representation of the inverse propagators.

\subsection{Calculating the determinants}

The operators $\mathcal{S}^{-1}$ and $\mathcal{S}_{(a)}^{-1}$ are 
defined in a $72$-dimensional vector space
$\Sigma=\mathcal{C}_{c,f}\otimes\mathcal{D}_{Dirac}\otimes\mathcal{N}_{NG}$. 
However, the charge projectors express $\Sigma$ as the direct sum of three 
different spaces,  i.e., $\Sigma=\Sigma_{(0)}\oplus\Sigma_{(+)}\oplus\Sigma_{(-)}$, 
with vector bases given by
\begin{equation}
\Sigma_{(0)}\Rightarrow\{|s_{1},\pm\rangle , |s_{2},\pm\rangle,|d_{1},
\pm\rangle,|d_{2},\pm\rangle,|u_{3},\pm\rangle \},
\label{basis0}
\end{equation}
\begin{equation}
\Sigma_{+}\Rightarrow\{|u_{1},+\rangle , |u_{2},+\rangle,|s_{3},-\rangle,|d_{3},-\rangle \},
\label{basisplus}
\end{equation}
\begin{equation}
\Sigma_{-}\Rightarrow\{|u_{1},-\rangle , |u_{2},-\rangle,|s_{3},+\rangle,|d_{3},+\rangle \},
\label{basisplus1}
\end{equation}
where $\{|+\rangle,|-\rangle\}$ is the basis of Nambu-Gorkov space. 
The color, flavor, and Dirac structures are automatically taken into 
account by describing a quark spinor $\psi$ as
\begin{equation}
\psi=\{s_{1},s_{2},s_{3},d_{1},d_{2},d_{3},u_{1},u_{2},u_{3}\},
\label{quarkspinor}
\end{equation}
where $(1,2,3)=(b,g,r)$ denotes the color indices. Thus, $\Sigma_{(0)}$, 
$\Sigma_{(+)}$, and  $\Sigma_{(-)}$ are vector spaces with dimensions 
$40$, $16$, and $16$, respectively.

We can now compute the determinant of the inverse propagator in 
Eq.~(\ref{Gamma}) in terms of its corresponding charge projections. 
In fact, one sees that the determinant splits into three separate pieces
\begin{equation}
\Gamma(T,\mu,\Delta,\phi,B)=\sum_{a=0,\pm}\Gamma_{(a)}=\sum_{a=0,\pm}\frac{1}{2}\ln \det\widetilde{\mathcal{S}}_{(a)}^{-1}.
\label{Gamma2}
\end{equation}
Note that in the evaluation of the determinants only the projection of $\mathcal{S}_{(a)}^{-1}$ on the corresponding lower dimensional subspaces $\Sigma_{(a)}$, which we call $\widetilde{\mathcal{S}}_{(a)}^{-1}$, is relevant. The inverse propagator $\widetilde{\mathcal{S}}_{(0)}^{-1}$ that appears in the evaluation of the determinant is defined as
\begin{equation}
\widetilde{\mathcal{S}}_{(0)}^{-1}=
\begin{pmatrix}
  [G_{0(0)}^{+}]^{-1}\otimes 1_{5\times 5}  &   \widetilde{\Phi}_{(0)}^{*}  \\
  \widetilde{\Phi}_{(0)} &  [G_{0(0)}^{-}]^{-1}\otimes 1_{5\times 5}      \\
\end{pmatrix},
\label{invprop0}
\end{equation}
whereas
\begin{equation}
\widetilde{\Phi}_{(0)}=i\gamma_{5}
\begin{pmatrix}
  0       &  0       &  0         &    \phi      &  \Delta    \\
  0       &  0       &   -\phi    &    0         &   0        \\
  0       &  -\phi   &     0      &    0         &   0        \\
  \phi    &  0       &     0      &    0         &  \Delta    \\
  \Delta  &  0       &     0      &    \Delta    &   0        \\
\end{pmatrix}.
\label{invprop0matrix}
\end{equation}
This matrix can be easily diagonalized and its eigenvalues are 
$|\phi|^2, |\phi|^2,|\phi|^2, \Delta_{1}^2 , \Delta_{2}^2$, where 
$\Delta_{1,2}^{2}=\frac{1}{4}(\sqrt{|\phi|^2+8|\Delta|^2}\pm |\phi|)^{2}$. 
Moreover, the other inverse propagators are given by
\begin{equation}
\widetilde{\mathcal{S}}_{(\pm)}^{-1}=
\begin{pmatrix}
  [G_{0(\pm)}^{+}]^{-1}\otimes 1_{2\times 2}  &   \widetilde{\Phi}_{(\pm)}^{*}  \\
  \widetilde{\Phi}_{(\pm)} &  [G_{0(\pm)}^{-}]^{-1}\otimes 1_{2\times 2}      \\
\end{pmatrix},
\label{invproppm}
\end{equation}
where
\begin{equation}
\widetilde{\Phi}_{(\pm)}=-i\gamma_{5}\Delta\,1_{2\times 2}.
\label{invproppmmatrix}
\end{equation}

Now, each term in Eq.~(\ref{Gamma2}) can be simplified using the identity in Eq.~(\ref{det}), which then gives
\begin{equation}
\Gamma_{(a)}=\frac{1}{2}\ln \det\left[[G_{0(a)}^{+}]^{-1}
[G_{0(a)}^{-}]^{-1}+\widetilde{\Phi}_{(a)}\widetilde{\Phi}_{(a)}^{*}\right],
\label{gammacharge}
\end{equation}
where we used that $[G_{0(a)}^{\pm}]^{-1}\widetilde{\Phi}_{(a)}^{+}
=-\widetilde{\Phi}_{(a)}^{+}[G_{0(a)}^{\pm}]^{-1}$. Since only the 
absolute square of the gaps appears in the expressions, from now on 
we take $\Delta,\,\phi >0$.

In order to compute the determinants it is convenient to use the chiral 
and energy projectors defined in Appendix B. First, it can be shown that
\begin{equation}
 [G_{0(a)}^{\pm}]^{-1}[G_{0(a)}^{\mp}]^{-1}
=\sum_{c=\pm}[\,p_{0}^{2}-(c|\bar{\mathbf{p}}_{(a)}|\pm \mu)^2]\Lambda_{(a)}^{c},
\label{twoinversepropagators}
\end{equation}
which then gives
\begin{widetext}
\begin{equation}
\Gamma_{(a)}=\frac{1}{2}\mbox{tr}_{c,f,\chi}
\sum_{p_0,\bar{\mathbf{p}}_{(a)}}\ln 
\left(\frac{p_{0}^{2}-(|\bar{\mathbf{p}}_{(a)}|-\mu)^2-\lambda_{(a)}^2}{T^2}\right)
+\frac{1}{2}\rm{tr}_{c,f,\chi}\sum_{p_0,\bar{\mathbf{p}}_{a}}
\ln \left(\frac{p_{0}^{2}-(|\bar{\mathbf{p}}_{(a)}|+\mu)^2-\lambda_{(a)}^2}{T^2}\right),
\label{gammacharge1}
\end{equation}
\end{widetext}
where $-\lambda_{(a)}^2$ are the eigenvalues of the matrices
$\widetilde{\Phi}_{(a)}\widetilde{\Phi}_{(a)}^{*}$. Also, $\mbox{tr}_{c,f,\chi}$
is the remaining trace over the color, flavor, and chiral indices. Moreover,
$|\bar{\mathbf{p}}_{(0)}|=\sqrt{p_{3}^2 + \vec{p}_{\perp}^{\,\,2}}$ and
$|\bar{\mathbf{p}}_{(\pm)}|=\sqrt{p_{3}^2 + 2eBn}$, where $n\geq 0$
is an integer that labels the Landau levels, and $\vec{p}_{\perp}$ is
the momentum perpendicular to the field. Also, it is clear now that
$\Gamma_{(+)}=\Gamma_{(-)}$. The sums in the equation above are defined as follows
\begin{equation}
\sum_{p_0,\bar{p}_{(0)}}f(p_0,\bar{\mathbf{p}}_{(0)})
=T\sum_{k}\int \frac{d^3\vec{p}}{(2\pi)^3}f(i\omega_{k},\vec{p}),
\label{sum0}
\end{equation}
and
\begin{equation}
\sum_{p_0,\bar{\mathbf{p}}_{(\pm)}}f(p_0,\bar{\mathbf{p}}_{(\pm)})
=T\sum_{k}\frac{eB}{8\pi^2}\sum_{n=0}^{\infty}\alpha_{n}
\int_{-\infty}^{\infty}dp_{3} f(i\omega_{k},n,p_{3}),
\label{sumpm}
\end{equation}
where $f$ is an arbitrary function, and $\alpha=2-\delta_{n0}$ stands for the fact that Landau 
levels with $n>0$ are doubly degenerate. Since these sums do not converge some sort of 
cutoff procedure has to be used. The Matsubara sum can be evaluated using the identity
\begin{equation}
 T\sum_{k}\ln\Big(\frac{\omega_{k}^2 +z^2}{T^2}\Big)=|z|+2T\,\ln(1+e^{-|z|/T}).
\label{matsubarasum}
\end{equation}

Therefore, we can now write the one-loop quark contribution to the pressure of the mCFL phase as
\begin{equation}
\Gamma=3 P_{T}(\phi)+P_{T}(\Delta_{1})+P_{T}(\Delta_{2})+4 F_{T}(\Delta).
\label{thermofunctionmCFLtemp}
\end{equation}
The function $P_{T}(\phi)$ is defined as follows
\begin{eqnarray}
\hspace*{-0.5in}P_{T}(\phi) &=& \mbox{tr}_{B=0}\left[E_{0}^{+}(\phi)+E_{0}^{-}(\phi)\right] \nonumber \\
&& \hspace*{-0.5in}+2T\mbox{tr}_{B=0}\left[\ln(1+e^{-E_{0}^{+}/T})+\ln(1+e^{-E_{0}^{-}/T})\right],
\label{functionPtemp}
\end{eqnarray}
where $\mbox{tr}_{B=0}[\ldots]=\int\frac{d^{3}\vec{p}}{(2\pi)^3}[\ldots]$,
$E_{0}^{\pm}[\phi]=\sqrt{(p \mp \mu)^2 +\phi^2}$, with
$p=\sqrt{p_{3}^2+\vec{p}_{\perp}^{\,2}}=|\vec{p}|$.
The zero temperature terms are divergent and they are regularized in Sec.~\ref{sectionII}.
The other function in Eq.~(\ref{thermofunctionmCFLtemp}) reads
\begin{eqnarray}
\hspace*{-0.5in} F_{T}(\Delta) &=& \mbox{tr}_{B} \left[E_{B}^{+}(\Delta)+E_{B}^{-}(\Delta)\right] \nonumber \\
&&  \hspace*{-0.5in}+2T \mbox{tr}_{B}\,\left[\ln(1+e^{-E_{B}^{+}/T})+\ln(1+e^{-E_{B}^{-}/T})\right]
\label{functionFtemp}
\end{eqnarray}
where $\mbox{tr}_{B}[\ldots]=\frac{eB}{8\pi^2}
\sum_{n=0}^{\infty}\alpha_{n}\int_{-\infty}^{\infty}dp_{3}[\ldots]$, 
$E_{B}^{\pm}[\Delta]=\sqrt{(\varepsilon_{B} \mp \mu)^2 +\Delta^2}$, 
with $\varepsilon_{B}=\sqrt{p_{3}^2+2eBn}$. Note that 
$\lim_{B\to 0}F_{T}(x)=P_{T}(x)$ and, once the limit $\Delta=\phi$ and 
$B=H=0$ is taken, we recover the free energy of the CFL phase in the 
absence of an external field.


\section{Properties of Ritus' eigenfunctions}
\label{appB}

The transformation functions $E_{P}^{\pm}$ for positively (+) and negatively (-) charged fermions are obtained as solutions of the field dependent eigenvalue equation
\begin{equation}
 (\Pi_{(\pm)}\cdot \gamma)E_{P}^{\pm}(X)=E_{P}^{\pm}(X)(\bar{P}_{(\pm)}\cdot \gamma),
\label{appendixAeigenvalue}
\end{equation}
where $\bar{P}_{(\pm)}=(p_0 , 0, \pm\sqrt{2eBn},p_{3})$ and
\begin{equation}
 E_{P}^{\pm}(X)=\sum_{\sigma}E_{P\sigma}^{\pm}(X)\Delta[\sigma],
\label{appendixA1}
\end{equation}
with eigenfunctions
\begin{equation}
 E_{P\sigma}^{\pm}(X)=\mathcal{C}_{\,l\,(\pm)}\,e^{-i(p_{0}x^0+p_2 x^2 + p_3 x^3)}\,D_{\,l\,(\pm)}[\rho_{(\pm)}],
\label{appendixAeigenfunctions}
\end{equation}
where $D_{\,l\,(\pm)}[\rho_{(\pm)}]$ are parabolic cylinder functions with argument $\rho_{(\pm)}$ defined as
\begin{equation}
 \rho_{(\pm)}=\sqrt{2eB}(x_1 +p_{2}/eB),
\label{appendixAdefinerho}
\end{equation}
and index $l\,(\pm)$ given by
\begin{equation}
 l\,(\pm)=n\pm\frac{\sigma}{2}-\frac{1}{2},\qquad
l(\pm)=0,1,2,\hdots,
\label{appendixAdefinelevel}
\end{equation}
whereas $n=0,1,2,\hdots,$ denotes the Landau levels and $\sigma$ is the spin projection that can take values $\pm 1$. Moreover, the normalization constant is
\begin{equation}
\mathcal{C}_{\,l\,(\pm)}=(4\pi\,eB)^{1/4}/\sqrt{l\,(\pm)\,!}\,.
\label{appendixAnormalization}
\end{equation}

The spin matrices $\Delta[\sigma]$ in Eq.~(\ref{appendixA1}) are 
spin projectors. They are defined as
\begin{equation}
\Delta[\sigma]={\rm diag}(\delta_{\sigma 1},\delta_{\sigma -1},\delta_{\sigma 1},\delta_{\sigma -1}),\qquad
\sigma=\pm 1,
\label{appendixADelta}
\end{equation}
and satisfy the following relations
\begin{subequations}
\begin{eqnarray}
\Delta[\pm]^{\dagger}=\Delta[\pm], &\qquad&
\Delta[+]+\Delta[-]=1,
\label{appendixADelta1}
\\
\Delta[\pm]\Delta[\pm]=\Delta[\pm], &\qquad&
\Delta[\mp]\Delta[\pm]=0,
\label{appendixADelta2}
\\
\gamma_{\parallel}\Delta[\pm]=\Delta[\pm]\gamma_{\parallel}, &\qquad& \gamma_{\perp}\Delta[\pm]=\Delta[\mp]\gamma_{\perp}.
\label{appendixADelta3}
\end{eqnarray}
\end{subequations}
In the equations above we used that 
$\gamma_{\parallel}=(\gamma^0 , \gamma^3)$ and $\gamma_{\perp}=(\gamma^1 , \gamma^2)$.

Massless quarks can be fully characterized by the chiral and energy projection 
operators,
\begin{eqnarray}
 P_{R,L} &=& \frac{1\pm \gamma_5}{2},
\label{appendixBchiral}
\\
\Lambda_{(a)}^{\pm} &=& \frac{1\pm \gamma_{0}\,\vec{\gamma}\cdot\hat{\bar{\mathbf{p}}}_{(a)}}{2},
\quad\mbox{with}\quad a=0,\pm.
\label{appendixBenergy}
\end{eqnarray}
respectively. Similarly to the free case, in the presence of a 
uniform magnetic field these two operators commute.

In Ref.~\cite{dirkpisarski} it was shown that only massless quarks with the 
same chirality pair in the spin zero channel. One can show that there are 
four different gap functions that describe the possible pairings. However, 
in an NJL theory with the gap functions independent of the 3-momentum, the 
total number of independent gaps is reduced to one. This means that the gap 
for quasi-particles and quasi-antiparticles are the same, in spite of the 
chirality of the particles. Moreover, the Dirac structure of the gap matrix 
is simply given by $C\gamma_{5}$, where $C=i\gamma^{2}\gamma^{0}$ is the 
charge conjugation matrix.

\end{document}